\documentclass[prl,bibnotes,twocolumn,showpacs, showkeys, superscriptaddress,preprintnumbers,amsmath,amssymb,lengthcheck]{revtex4-2}

\usepackage{amsmath}
\usepackage{graphicx}
\usepackage{dcolumn}
\usepackage{bm}
\usepackage{hyperref}
\usepackage{xcolor}

\newcommand{\fig}[1]{Figure \ref{#1}}

\newcommand{\Eq}[1]{Eq. \ref{#1}}

\newcommand{\Citeappendix}[1]{Appendix \ref{#1}}

\begin{document}

\title{Superlocalization reveals long-range synchronization of vibrating soliton molecules}

\author{Said Hamdi, Aur\'{e}lien Coillet, Benoit Cluzel, Philippe Grelu,and Pierre Colman $^{*}$ }

\affiliation{Laboratoire Interdisciplinaire Carnot de Bourgogne UMR CNRS 6303, Universit\'{e} Bourgogne--Franche-Comt\'{e}, 9 av. Savary, 21000 Dijon, France}
\email{$^*$  Corresponding author: pierre.colman@u-bourgogne.fr } 

\date{\today}

\begin{abstract}
We implement a super-localization method in the time domain that allows the observation of the external motion of soliton molecules in a fiber ring cavity laser with unprecedented accuracy. In particular, we demonstrate the synchronization of two oscillating soliton molecules separated by several nanoseconds, with inter-molecules oscillations following the same pattern as the intra-molecular motion of the individual molecules. These experimental findings indicate an interplay between the different interaction mechanisms that coexist inside the laser cavity, despite their very different characteristic ranges, timescales, strengths, and physical origins. 
\end{abstract}

\keywords{Point Spread Function Deconvolution, Soliton Molecule, HyperResolution, Long range interaction, Multiscale Interplay}
\maketitle

\section{\label{sec:Intro}{Introduction}}

\par In recent years, the investigation of the dynamics of optical soliton molecules in ultrafast lasers has attracted increasing attention \cite{Grelu2008}. Soliton molecules consist in solitons forming a bound state owing to their relatively strong interaction through the propagation medium \cite{Stratmann2005}. Sometimes, these bound states exhibit a periodic internal motion that bears analogy with the vibration of molecules in chemistry. Previous real-time experiments highlighted different oscillatory behaviors, such as vibrations, pure phase, or anharmonic oscillations \cite{Krupa2017,Herink2017,Hamdi2018}. The specific oscillation pattern depends on the laser parameters, which determine the existence and the intrinsic properties of a multi-dimensional limit-cycle attractor for dissipative solitons \cite{Grelu2012}. The actual dynamics is also affected by laser noise \cite{Weill2016} and other experimental perturbations \cite{Nimmesgern21,Zhou2021,He2019}. Beyond their fundamental appeal, soliton molecules could be involved in multi-pulse patterns of practical interest, such as in harmonic mode-locking or in optical data manipulation \cite{Akhmediev2001}. Thus, experiments investigate now soliton molecule dynamics for a larger number of interacting solitons. In a significant number of cases, solitons are not distributed evenly in a regular train of pulses, but form a supra-molecule or molecular complex composed of several, often identical, molecules \cite{Wang2019, Luo:20, He2019}. This raises the open question of the interplay between the mechanisms responsible for the dynamics governing each molecule, and for the macro organization of the molecular complexes. 

Many experiments regarding soliton molecules rely on the dispersive Fourier-transform (DFT) technique \cite{Goda2013}, which allows a single-shot real-time recording of the spectra over successive cavity roundtrips at multi-MHz frame rates. After numerical processing, DFT spectra yield the dynamics of relative timing and phase between temporally separated pulses. As a major limitation, the related observation windows $T_\text{obs}$ is limited to pulse separations typically below a hundred of picoseconds \cite{Nimmesgern21,Zhou2021,He2019,Wang2019}. Such practical limitation is due to the finite spectral resolution, which is bound on one side by the speed of the detection electronics and on another side by the magnitude of the dispersive line used for pulse stretching. Indeed, the need to avoid the overlap of the DFT traces for two subsequent pulses prevents the use of a dispersive line of arbitrary length. To set typical orders of magnitudes, considering a electronic bandwidth of $\sim$ 10 GHz and a repetition rate of $\sim$ 10 MHz for $\sim$ 200 fs pulses yields approximatively $T_\text{obs} \sim$ 100 ps. Moreover, whereas the DFT recording can retrieve the internal motion of a given soliton molecule, it does not provide an accurate information regarding the global motion of the molecule round the laser cavity. Instead, the direct observation - i.e. without pulse stretching - using a GHz-bandwidth oscilloscope can provide a temporal resolution down to a few tens of picoseconds. Nevertheless, such timing precision is not enough to observe the weak timing fluctuations that are expected to take place within the supra-molecule. Indeed the strongest mechanisms have the shortest range of interaction\cite{Grelu2008}, while long range interactions are much weaker \cite{Jang2013,Pilipetskii1995,Weill2016}. Consequently observing the latter would require an acute measurements precision, about on par with the resolution provided by the DFT, but with a much larger observation windows.

Thus, we have complemented the DFT with a direct timing channel to record the time of passage of the soliton molecules. And in order to improve the accuracy of the long-range timing, we have implemented on these timing traces a super-localization procedure akin to what is done in fluorescence spectroscopy \cite{Patterson2009,MICHALET2001,Thompson2002}. This way, we got a timing resolution down to about 0.14~ps -- 100 times better than the sampling resolution -- allowing us to precisely correlate the internal vibrations of the soliton molecule with its external motion in the moving frame. Subtle changes of the molecules cruising velocity round the laser cavity can now be observed.

We applied this new technique to the co-evolution of two soliton-pair molecules \cite{Hamdi2018}, which are separated by $\simeq$ 7.6~ns, one third of the roundtrip time (23.0 ns, $\text{FSR}=43.4$~MHz). For a clear demonstration of the possibilities offered by combining short and long time scales measurement, two soliton-pair molecules make indeed the simplest multi-molecular system. 


\section{\label{sec:IntMotion}{Internal Motion: Twin Molecules} }
\begin{figure}[htbp]
	\centering
	\includegraphics[width=1\linewidth]{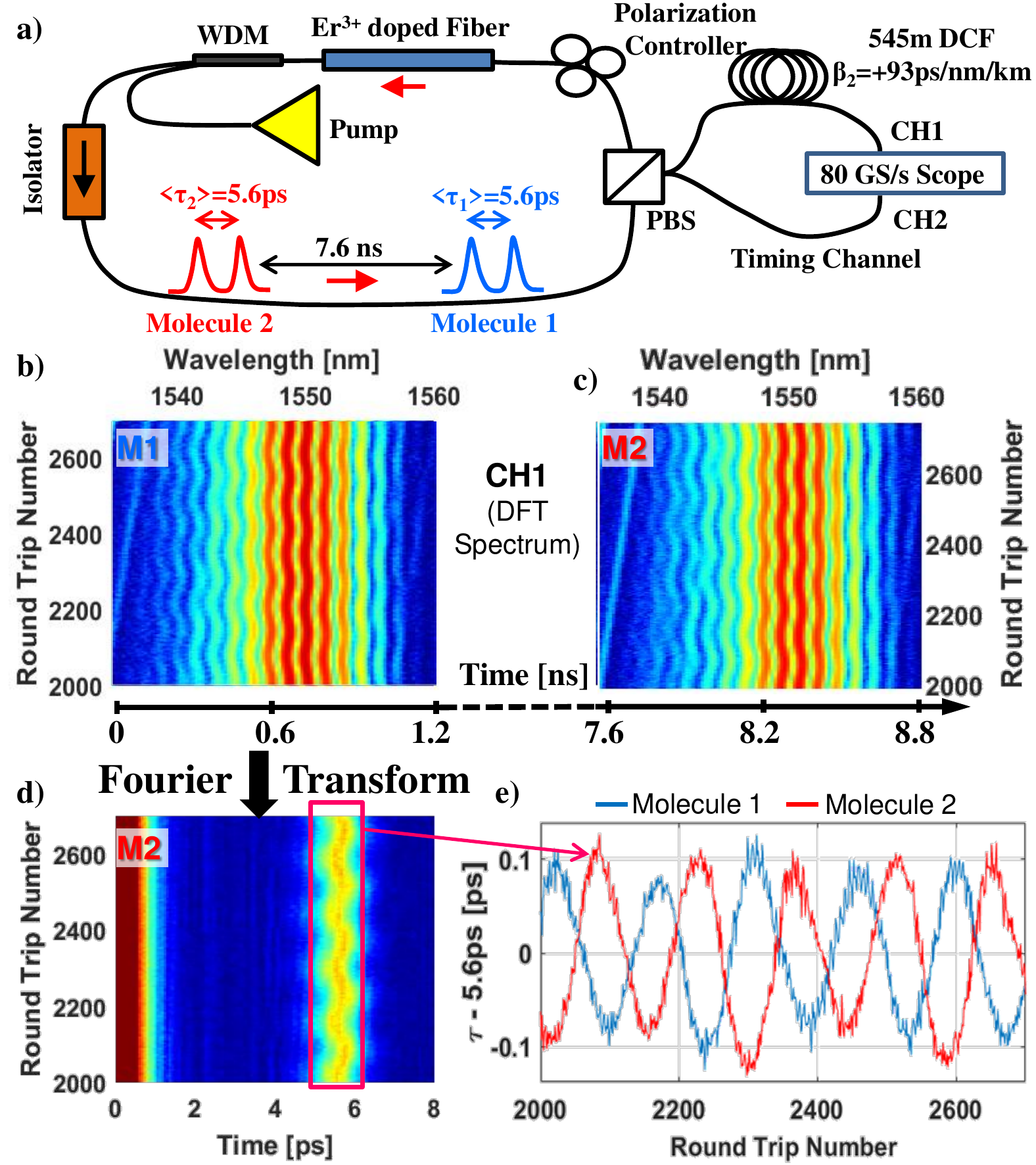}
	\caption{\label{Fig:1} a) Experimental setup; PBS: polarization beam splitter; DCF: 545~m of dispersion-compensating fiber (-90.15~ps/nm/km); Pump: 980-nm laser diode; WDM: pump-signal multiplexer. Single-shot spectra in color scale for b) the leading and c) the trailing soliton molecule. d) Temporal autocorrelation traces after Fourier-transform of c). e) Internal vibration motion for each molecule. Blue (resp. red) line stands for the leading (resp. trailing) molecule. They both exhibits an average soliton separation of 5.57~ps.}
\end{figure}

The fiber ring laser cavity under study is composed of 1 meter of erbium-doped fiber closed by 3 meters of SMF fiber (see \fig{Fig:1}-a)). Nonlinear polarization evolution (NPE) in the fibers provides a virtual saturable absorber effect for mode locking: a polarization beam-splitter (PBS) converts NPE into amplitude modulation and provides as well a laser output port. The first channel of a 80~GSamples/s oscilloscope records the DFT spectrum after an accumulated a chromatic dispersion of -49~ps/nm. This channel yields the information about the internal motion for each molecule. The second channel records the direct pulsed laser output, which is used to track down the global motion of the molecules round the cavity. A precise measurement of the delay between both input channels allows the unambiguous attribution of a DFT spectrum to each pulse observed on the timing trace. The direct detection reveals the presence of two sets of pulses and the DFT signal confirms that both sets are actually stable pairs of solitons separated by $5.57\pm 0.01$~ps. 

The internal vibration of each soliton molecule is obtained by Fourier transform of the corresponding DFT signal, as illutrated by \fig{Fig:1}-b,c,d). We see in \fig{Fig:1}-e) that the two molecules both follow a common periodic oscillation with an amplitude of $92 \pm 6$~fs and a periodicity of 143.5 cavity round trips (RTs). The maximal correlation coefficient between the two trajectories is 0.89 for a delay of 87.2~RTs, which corresponds to a phase-shift of $219^{\circ}$. This strong correlation means that the two oscillations belong to the same family: about the same amplitude and periodicity and fluctuations, minus some slight phase drifts - as discussed later in this article. Since the existence of soliton molecules and their vibration patterns are fixed by a common dissipative attractor \cite{Grelu2012}, it is not surprising that the dynamics of the two pairs exhibit nearly the same features. However, the two molecules are not isolated physical objects. Instead, they are likely to interact, as the existence of weak long-range (ns) interactions has been established \cite{kutz98,Pilipetskii1995, Weill2016}. Therefore, one could wonder whether the two molecules following a common vibration pattern would benefit from an additional synchronization mechanism. In theory, several distinct vibration patterns could have coexisted \cite{Schreiber2007,Zavyalov2009a,Zavyalov2009}.

\section{\label{sec:ExtMotion}{External Motion: Point Spread Function (PSF) deconvolution} }

\begin{figure}[htbp]
	\centering
	\includegraphics[width=1\linewidth]{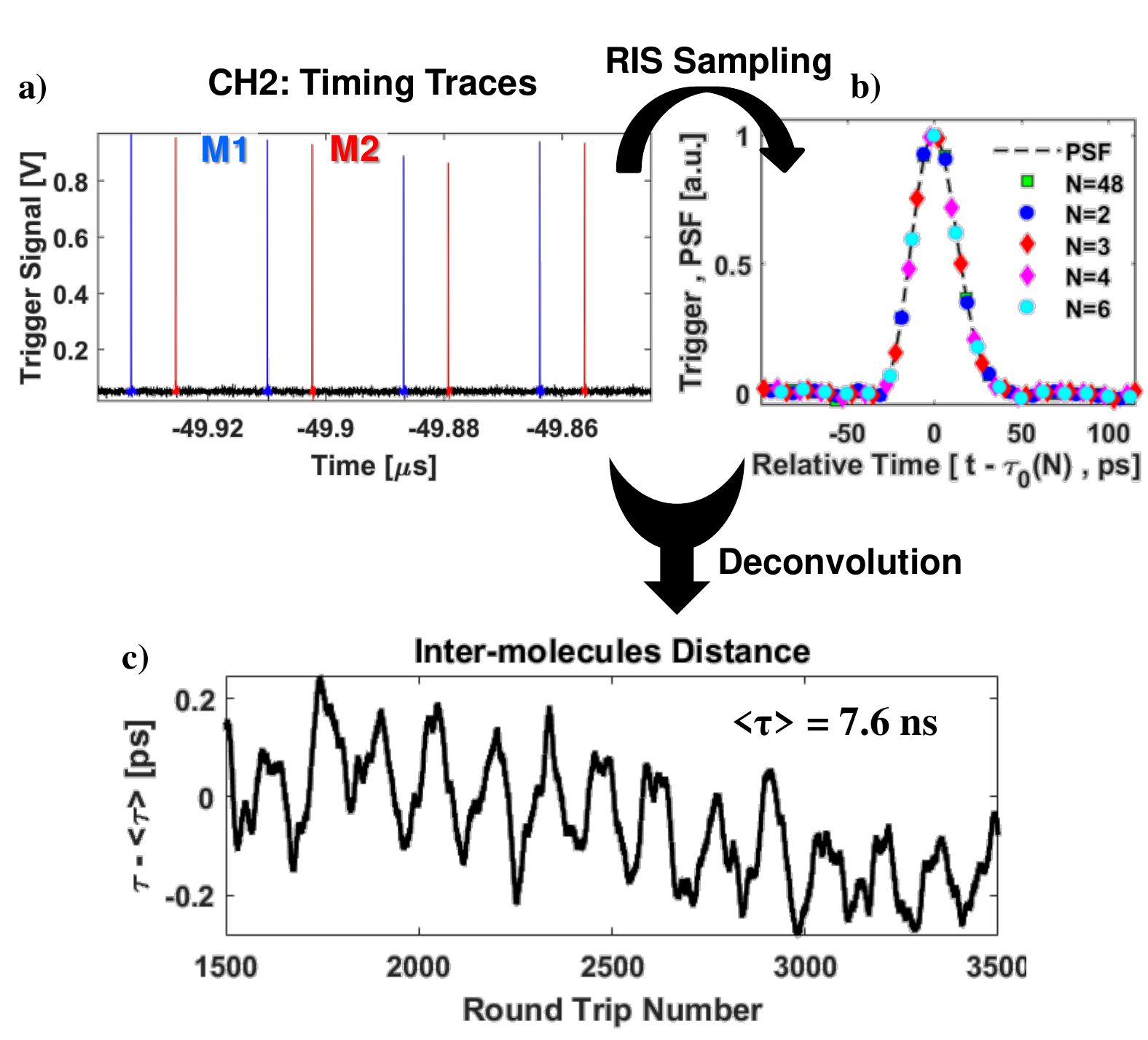}
        \caption{\label{Fig:2} a) Timing trace displaying one intensity peak per molecule and per round-trip (RT). b) Reconstruction of the point spread function (PSF) for M1: the random interleaved sampling (RIS) effect between the cavity RT time and the oscilloscope's sampling clock is shown for a few timing pulses. Dashed-black: reconstructed PSF after averaging over 4221~RTs. c) Evolution of the temporal separation between the two molecules (M1, M2) with sub 20-fs resolution. The average inter-molecular separation is $<\tau_{\text{ext}0}>~=$ 7.58~ns.}
\end{figure}

To investigate this matter, we examined further the timing pulse signal. In order to improve the native 12.5~ps (80~GS/s) sampling resolution, we super-localize the pulses using the point spread response function (PSF) of the detection link. Such a deconvolution technique is widely used in fluorescence microscopy to achieve spatial super-resolution \cite{Patterson2009,MICHALET2001,Thompson2002}. In a nutshell, by superposing together all the pulses acquired at each round-trip, it is possible to take advantage of the aliasing between the cavity free spectral range and the oscilloscope's internal sampling clock to perform a random interleaved sampling (RIS) and retrieve the shape of PSF with at least a tenfold improvement in the temporal resolution (\fig{Fig:2}-b) ). The large number of timing traces (4221~RTs here) also allows averaging the timing traces, reducing further the noise. Once the high resolution PSF is obtained, it is fitted back on each timing pulse to determine precisely its time of passage. By doing so, we determine the times of passage for each molecule with a precision down to 140~fs. This residual uncertainty is due to a mixture between the laser intrinsic jitter, and the jitter of the oscilloscope' sampling clock. More details regarding the implementation of the PSF deconvolution can be found in \Citeappendix{App:A}. A major difference with fluorescence microscopy is that the signal measured here is much stronger, and more events are also recorded, resulting in a better improvement. 

The advance or delay for each molecule compared to the average round trip time is shown in \fig{Fig:2}-c). The most striking feature is that the distance between the two molecules oscillates, following a pattern very similar to the internal motion: same periodicity and nearly the same amplitude ($106 \pm 18$~fs versus $92 \pm 6$~fs). Similarly to the internal motions, this external oscillation has its own phase offset.

\begin{figure}[htbp]
	\centering
	\includegraphics[width=1\linewidth]{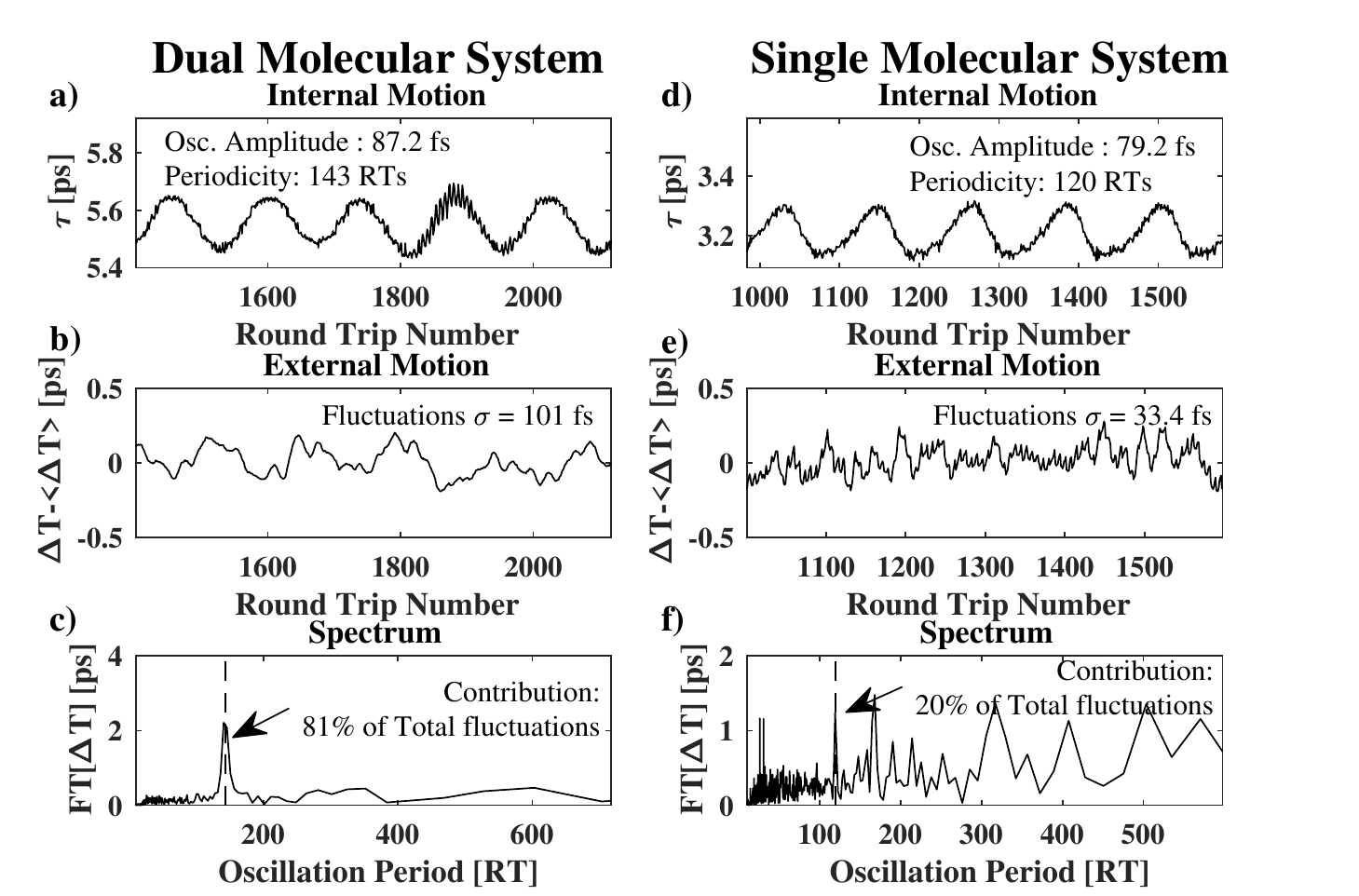}
        \caption{\label{Fig:3} a) Internal vibrational motion within the soliton molecule M1 composing the two-molecule system. b) Variation of the inter-molecular distance with respect to the round trip number. c) Fourier transform of b): the dashed line indicates the period of the internal vibration shown in a). d-f) Same as a-c, but typical results for a single-molecule system.}
\end{figure}

In order to better understand the specificity of this inter-molecular oscillation, we compared in \fig{Fig:3} the dynamics of the dual-molecule system with that of a single-molecule, recorded from the same laser setup. Both dynamics show nearly the same internal oscillatory motion (\fig{Fig:3}-a,d)). However, the two-molecule system exhibits higher fluctuations of its external motion relative to the moving propagation frame (\fig{Fig:3}-b,e)). Using the Parseval's theorem, we can estimate the amount to which a peculiar periodicity contributes to the external relative motion. For the two-molecule system, most of the external motion (81\%) can be described as an oscillation with the same periodicity as the internal motion. There are no other notable contributions. On the contrary for the single molecule system, the trajectory around the cavity is mainly composed of noise fluctuations which are not related to the internal vibration. The remaining 20\% contribution synchronized to the molecule's internal oscillation remain hypothetical: they could be inferred as a small influence of the molecule dynamics on its velocity, or the result of a periodic change in the center of mass of the molecule which would be caused by a strong asymmetry in the vibration (e.g. a significant exchange of energy between the two solitons).

\section{\label{sec:Interaction} Long-range interaction and synchronization}

Correlation does not imply causation; however, clues regarding any interaction between the different oscillators can be inferred by examining how the oscillators respond to fluctuations. Indeed, since the oscillators are separated by several nanoseconds, they do not experience exactly the same random noise. Hence, fully independent and non-interacting oscillators would slowly drift from each other following a random walk pattern. By examining how the oscillators stay correlated, and the dynamics by which they drift away, one could further infer whether (i) the oscillators are phase-locked to each other, (ii) they are synchronized, hence their respective evolution show slight differences but some back-action mechanism prevents them from drifting beyond a certain range from each other, (iii) they evolve completely independently.

\begin{figure}[htbp]
	\centering
	\includegraphics[width=1\linewidth]{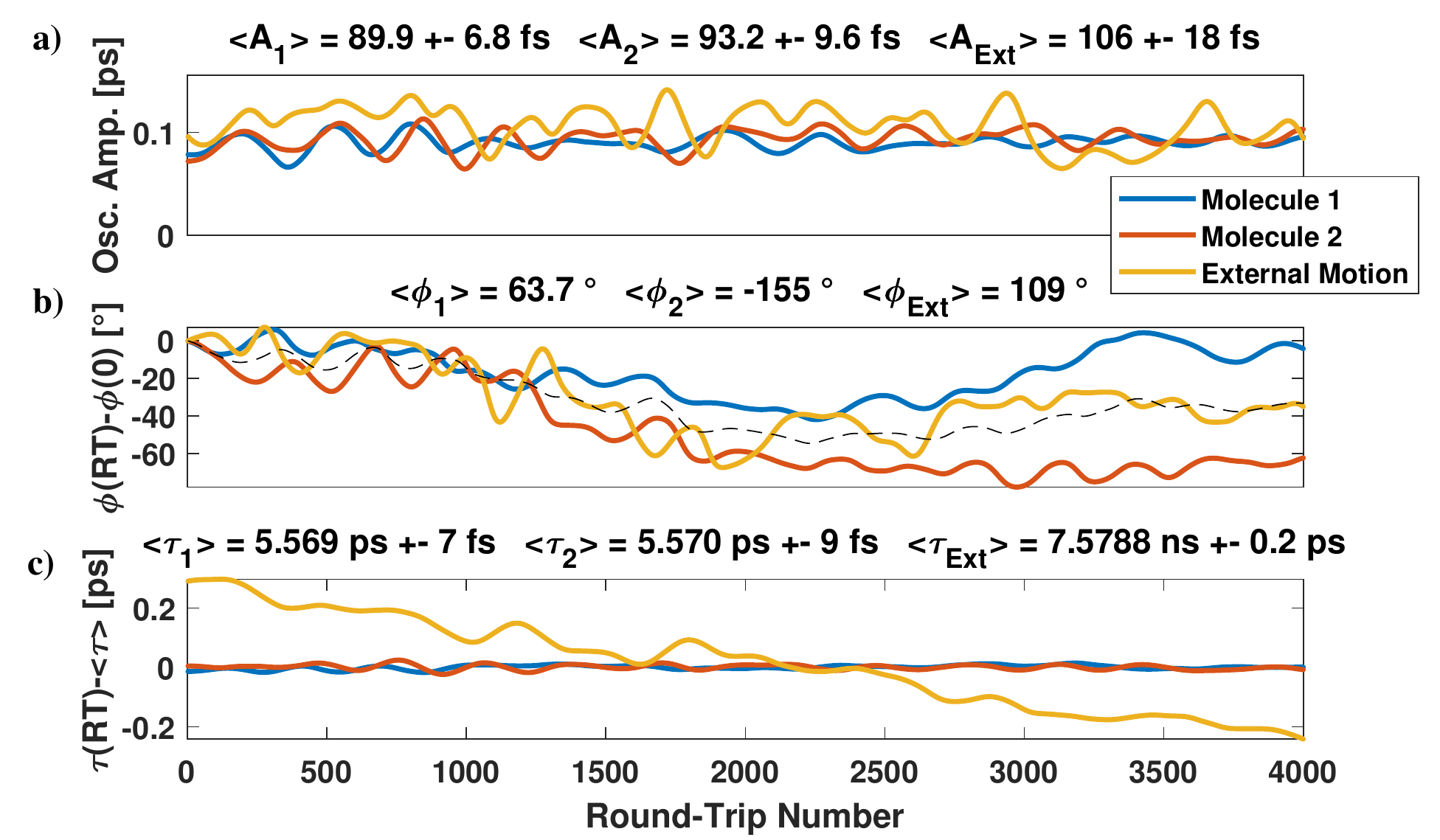}
	\caption{\label{Fig:4} a) Instantaneous oscillation amplitude for molecule M1 ($A_1$), M2 ($A_2$), and for the inter molecular oscillation ($A_{ext}$). b) Similar to a) but regarding the relative phase of the oscillation. Black-dashed : result of eq.(1). c) Evolution of the average soliton separation for M1 (blue) and M2 (red), and of the distance between the two molecules (yellow).}
\end{figure}

For slow and weak fluctuations (adiabatic regime), the motion $\tau(t)$ of a harmonic oscillator around its equilibrium position $\tau_0$ can be described as $\tau (t) = \tau_0(t) + A(t) \cos(\omega t + \phi(t))$, where $\tau_0$, $A$, and $\phi$ are the equilibrium position, the oscillation amplitude, and the phase offset, respectively. These parameters fluctuate under the action of noise and drift of the lasing conditions. Therefore, it is possible to get a finer description of the dynamics of the oscillators by monitoring the evolution of the latter three parameters, as shown in \fig{Fig:4}.

Firstly, the oscillation amplitudes for the internal motions are more stable than for the external relative motion \fig{Fig:4}-a). Noting the small temporal extension of each soliton molecule ($<7$~ps here), we can interpret the moderate amplitude excursions as resulting from a stronger stabilizing interaction between the two optical solitons that compose each molecule. Indeed, these two dissipative solitons are subjected to phase-sensitive short-range interactions \cite{Malomed1991,Soto-Crespo2003}. This strong binding attenuates the impact of noise and the drifting associated to it. In addition, the impact of external noise depends on the separation between the oscillator's constituents. Indeed, for a given temporal separation $T$, the oscillator will be mainly affected by noise components above the cutoff frequency $1/T$ since these high-frequency components will affect differently the pulses. Therefore, the larger the separation, the larger the noise bandwidth affecting the oscillator's dynamics. The weaker interaction and greater sensitivity to low frequency noise explain the faster drift observed for the external oscillator, as displayed on \fig{Fig:4}-c. That said, considering its spatial extension, the external oscillator has a relative drift proportionally much smaller than that of the oscillating molecules. 

Regarding the link between the two molecules, hence the external motion, long-range (nanosecond) interactions are very difficult to model precisely in fiber lasers, because they can combine physical effects such as gain depletion and recovery, electrostriction, and random walk on a noise floor \cite{kutz98,Jang2013,Pilipetskii1995,Weill2016}. In contrast with short range interactions, long range effects are much weaker. In addition, they are sensitive to the optical intensity of the pulses but not to the relative phase between the dissipative solitons. For close-by solitons, the relative optical dephasing is an important features that controls the attraction and repulsion, hence the oscillation of the molecule. A priori there does not exist any mechanism that would control the relative optical phase offset between the very distant two molecules. However a crucial point here is that the internal motion of the molecule is an oscillator that possesses its own phase, which is of different nature than the optical phase. Therefore, it is still possible that long-range interactions become sensitive to the phase of the soliton molecule oscillators. Some indications about the possible synchronization between the three oscillators, can be found in \fig{Fig:4}-b) that features the evolution of the relative phase offset $\phi(RT)- \phi(0)$. As the two molecules circulate round the cavity, their respective internal motions experience different phase fluctuations and drift away significantly from each other. When it comes to the external motion, we can interpret that its phase evolution is similar to the one that would experience a simple spring-like oscillator when excited at both ends by two oscillatory forces that are not phase-locked to each other. Basically, the phase evolution of such a spring oscillator would satisfy:
\begin{gather}
    \cos(\omega\,t+\phi_{spring}(t)+\delta) = \\ \nonumber 
    \cos(\omega\,t+\phi_1(t)) + \cos(\omega\,t+\phi_2(t)) 
    \label{eq_motion}
\end{gather}
Where $\phi_1(t)$ and $\phi_2(t)$ are the relative phases of the two driving forces, which we assume here to be the relative phases of the internal motion for each molecule. $\delta$ is a constant phase offset that accounts for a delayed response of the oscillator to the external driving, taken here as $\delta$ $\simeq$ $20^{\circ}$ (equivalent to an offset of 8 RTs), which yields the dashed line in \fig{Fig:4}-b. Therefore, the present data analysis indicates that the external oscillation is most likely driven by the combination of the internal motions of the soliton molecules. 

\section{\label{sec:Conclusion} Discussion and conclusion}
The above analysis is limited by the number of recorded rountrips (4221 RTs), itself limited by the oscilloscope memory, so that we cannot give a definite statement regarding the strength of the synchronization between the internal and external motions of the soliton molecules. Nevertheless, the progress of our argument in that direction is made possible by the super-localization techniques and refined data analysis which shed new lights on the complex nonlinear dynamics of multiple soliton molecules. Therefore, by presenting such tools and approach, our article should stimulate further experiments aimed at exploring the dynamics of soliton molecular complexes with an acute precision. We have also analyzed a second experimental data set, namely consisting in another set of oscillating soliton-pair molecules, which is presented in \Citeappendix{App:B}). Endowed with similar but not identical features, it confirms the existence of an external oscillation relative to the separation between the soliton molecules. It also leaves the possibility for a synchronization between the external oscillator and the internal ones. 

Let us elaborate on the theoretical possibility that two widely separated oscillating soliton molecules would set off an oscillation of their relative distances. The distance between the molecules refers to the distance between their centers of mass. There are two main possibilities. The first possibility is that the external oscillation do not result from long-range interaction, but rather from the local oscillation of the center of mass the molecule, which is caused by a slight symmetry breaking among the two pulses that constitute a soliton molecule \cite{Tchofo19, COLMAN_CLEO21}. Indeed, a vibrating soliton molecule is not a simple two-parameter oscillator but is truly multidimensional \cite{COLMAN_CLEO21}, with in particular energy flow occurring within the molecule. However, our experimental analysis of a single molecular system, presented in \fig{Fig:3}, rules out this first possibility. \\The remaining possibility is that the external oscillation results from a long-range interaction that is sensitive to the phase of the intra-molecular motion. In turn if long-range interactions are not sensitive to the relative phase between the electric fields envelopes of the pulses, the short range ones are, hence the intra-molecular state of vibration. Besides, the optical intensities of the pulses can also slightly oscillate, which would favor long range interplay that depends solely on the pulses energy.
	
At this point, it is essential to discuss into more details the soliton molecules dynamics. Occurrences of multiples soliton molecules systems, or of supra-molecular assembly, have been previously reported in the literature \cite{Wang2019,Zhou2020}. Studies focus either on the binding mechanisms controlling the equilibrium position between the solitons/molecules \cite{He2019, Zhou2020, Zhou2021}; or can focus more on the internal molecules' dynamics and their possible interaction \cite{Wang2019}. A second distinction concerns the type of interactions that are involved. Indeed they can be sensitive or not to the optical phase of the solitons, hence have different repulsive and attractive landscapes. Therefore depending on the interaction that are involved, completely different dynamics are to be expected. In a nutshell, multi-molecules systems are characterized by at least two main features: which effect is responsible for the macroscopic organization of the molecules? Which effect controls the individual properties -hence vibration pattern- of each molecules? And is there any mutual interplay between the macroscopic and microscopic arrangements of the soliton? It is then also equally important to distinguish between on one side the adiabatic adaptation of the equilibrium positions to an external stimuli related to the change of the laser parameters \cite{Zhou2021}; and on the other side the intrinsic molecules dynamics that occurs at (about) constant laser parameters. Only the latter can be considered as a true oscillator. Thus far, the intrinsic limit of the DFT technics, as discussed in the introduction, limited the studies to situation where the supramolacular cohesion is governed by the same interactions as for the individual molecules binding: short range interactions. Moreover these are strong coherent interactions. Thus, this results in a natural and strong interplay between the internal and external oscillators; and leads to a perfect locking of the molecules \cite{Wang2019}. In this article, the long range organization of the molecules, which are separated by precisely one third of cavity round trip, is well ascribed by acoustic and optomechanical interactions \cite{Erkintalo15,Jaouen2001}. These interactions are of different nature than a direct soliton interaction as they are not sensitive to the optical phase. Nevertheless we showed that they may be sensitive to the oscillator's phase and create thus a link between the internal motions. This synchronization effect is very weak, in particular with regard to the laser noise, but it can accumulate over numerous round trips \cite{Jaouen2001}; and hence lead to noticeable effects: the two molecules have the same oscillation pattern. The direct observation of the said synchronization mechanism requires however an extremely precise timing with sub pico-second resolution.
\medskip
\par To conclude, we have shown that soliton molecules influence each other over long-range. Despite being separated by several nano-seconds, compact picosecond soliton molecules cannot be considered as independent. Interestingly, long range interactions do not only determine the relative position of the molecules, but can also serve as media to exchange dynamical information by coupling the internal degree of the molecules together. Despite the fact that such possible synchronization will remain weak and cannot give rise to a strict locking, it is an open question whether  different molecules could share the same vibration properties and can thus drive each other resonantly. Obviously, the mechanisms involved here would also be present in harmonically mode-locked fiber lasers \cite{Liu2019, He2019}. 

The Point Spread Function deconvolution that we implemented is critical in order to unveil such subtle interactions, as this requires a very precise timing resolution below 300~fs. This technique provides new possibilities to study the dynamics of multi-pulse systems in fiber ring cavity lasers \cite{He2019}. As a pure numerical post-measurement processing, this deconvolution technique is easy to implement. Note that the final timing resolution after deconvolution depends on the quality of the experimental implementation. As a practical example, we managed in this paper a two order of magnitude improvement of the resolution.

Concerning the specific topic of soliton molecule analysis, performing an instantaneous phase and amplitude analysis – as shown in \fig{Fig:4} – is essential to describe with accuracy the vibration patterns. In particular it provides information about the fluctuations, the latter being directly related to the laser noise. Therefore,  fluctuations of the molecule vibration reveal the intrinsic properties of the limit-cycle attractor in response to the environment noise. Thus far, description of soliton molecules has been mostly qualitative, with scarce information regarding the stability of the molecules' motion. We hope the present work will serve as incentive for more thorough and quantitative analysis of soliton molecules' vibration patterns.

	\medskip
 This work was supported by the French “Investissements d'Avenir” program / project ISITE-BFC (contract ANR-15-IDEX-0003), and by the Agence Nationale de la Recherche (ANR) project ``CoMuSim'' (contract ANR-17-CE24-0010-01). 
	
	\medskip
	
	The authors declare no conflicts of interest.
	


	

\appendix
\section{\label{App:A} Determination of the Point Spread Function; and deconvolution procedure}
The precise determination of the time of passage is limited by three factors: \textit{(i)} the finite sampling rate of the oscilloscope, \textit{(ii)} the measurement noise, and \textit{(iii)} the presence of a background noise. In the paper, the first factor limits \textit{a priori} the temporal resolution to 12.5~ps (sampling rate is 80~GS/s). The simplest approach to improve the resolution would be to define the time of passage as a weighted momentum of the whole timing trace. However, the use of a first order momentum is intrinsically limited by the fluctuations of background noise because the timing pulse is short - its spans over a few sampling points, between 2 and 4 - so that the numerous extra data points that constitute the background have a great influence. In turn, an higher order weighted momentum dramatically reduces the background's contribution, but increases the impact of measurement noise. In contrast, the matching of the timing trace with regards to a well determined Point Spread Function can remove both background fluctuation and measurement noise. 

The fitting of the timing pulse against the PSF is a simple task as it is just a matter of translating and scaling the nominal PSF so that it best matches the few data points that constitute the timing trace. The main difficulty is to determine the PSF with enough accuracy.

\subsection{PSF determination and metrics}

Providing one knows with enough accuracy the time of passage $\tau_0(N)$ of each molecule for each round trip $N$, it is then possible to superpose all the timing trace on top of each other. This operation is equivalent to performing a random interleaved sampling (RIS) because the sampling rate of the oscilloscope and the cavity round trip time are not matched. Moreover, once a new timing resolution is chosen, all the data-points belonging to the same time-bin can be averaged together to reduce the measurement noise. In details we set the amplitude $Y(t)$ of the PSF at a given time $t$ as the result of the weighted least square error (W-LSE) fitting of a polynom of 4 order over a time windows of 4~ps span centered around $t$. Considering the overall much slow response of the detection link, the choice of the polynom's order and of the windows' width have little influence on the final result. The weight that is set on each data point is determined by the error given by a prior non-weighted LSE fitting. The weighted fitting has the advantage to decreases the impact of spurious points and oddly behaving timing pulses. All in all, the PSF is defined with a 10-fold temporal resolution and about 4-bit $Y$ resolution improvement compared to the oscilloscope's specifications. Note that the temporal random interleaved sampling (RIS) has no equivalent in optical microscopy where only noise reduction is possible. As a result the increase of super-localization are smaller than what we achieve here. There is clearly a trade-off between the increase in the temporal resolution and the reduction of the $Y$ measurement noise. Thus far we have not checked what would be the best trade-off.

If knowledge of the PSF allows the precise determination of $\tau_0(N)$, a (good) estimation of $\tau_0(N)$ is at the same time required in order to reconstruct the PSF. The PSF and $\tau_0(N)$ must then be defined using a self-consistent procedure. The measured $\tau_0^{\text{Meas}}(N)$ is composed of several elements :

\begin{align}
  \tau_0^\text{Meas}(N) &=  N~\tau_{RT} + \delta^\text{Meas}(N) \\
    &=  N~\tau_{RT} + \delta^\text{slow}(N) + \delta^\text{p2p}(N) + \delta^\text{error}(N) \nonumber
  \label{eq_error}
\end{align}
$\tau_{RT}$ is the average round trip time, $N$ the round trip number, while the other terms represent the variations of the time of passage with respect to a perfectly periodic circulation. $\delta^\text{slow}(N)$, $\delta^\text{p2p}(N)$, and $\delta^\text{error}(N)$ correspond respectively to the slow (about 143.5 RTs period here) evolution of the molecule around the laser cavity, the pulse to pulse jitter, and the timing error that we make. The standard deviation of $\delta^\text{Meas}$ taken over all the round trips ($n=\infty$) is then 
\begin{equation}
  \sigma^2_\text{Meas}(n) = \sigma^2_\text{slow}(n) + \sigma^2_\text{p2p}(n) + \sigma^2_\text{error}(n)
  \label{eq_dev}
\end{equation}
$\sigma^2_\text{slow}(n=\infty)$ dominates over $\sigma^2_\text{p2p}(n=\infty)$. However if we consider a number of $n$ consecutive round trips much shorter that characteristic slow evolution of the soliton molecule then $\sigma^2_\text{slow}(n) \approx 0$. Considering the molecules oscillate over 143.5 RTs, we chose $n=8$. The expected value of the standard deviation over all the possible $n$ consecutive bin, reads then:
\begin{align}
  <\sigma^2_\text{Meas}(8)> & \approx <\sigma^2_\text{p2p}(8)> + <\sigma^2_\text{error}(8)> \nonumber \\
      & \approx C^{te} + <\sigma^2_\text{error}(8)>
  \label{eq_dev_final}
\end{align}
We define \Eq{eq_dev_final} as our metric. It is minimal when $\delta^\text{Error}$ is minimal - ideally null.

\subsection{First pass : Initial evaluation of $\tau_0^\text{Meas}(N)$}
\begin{figure}[htbp]
  \centering
  \includegraphics[width=1\linewidth]{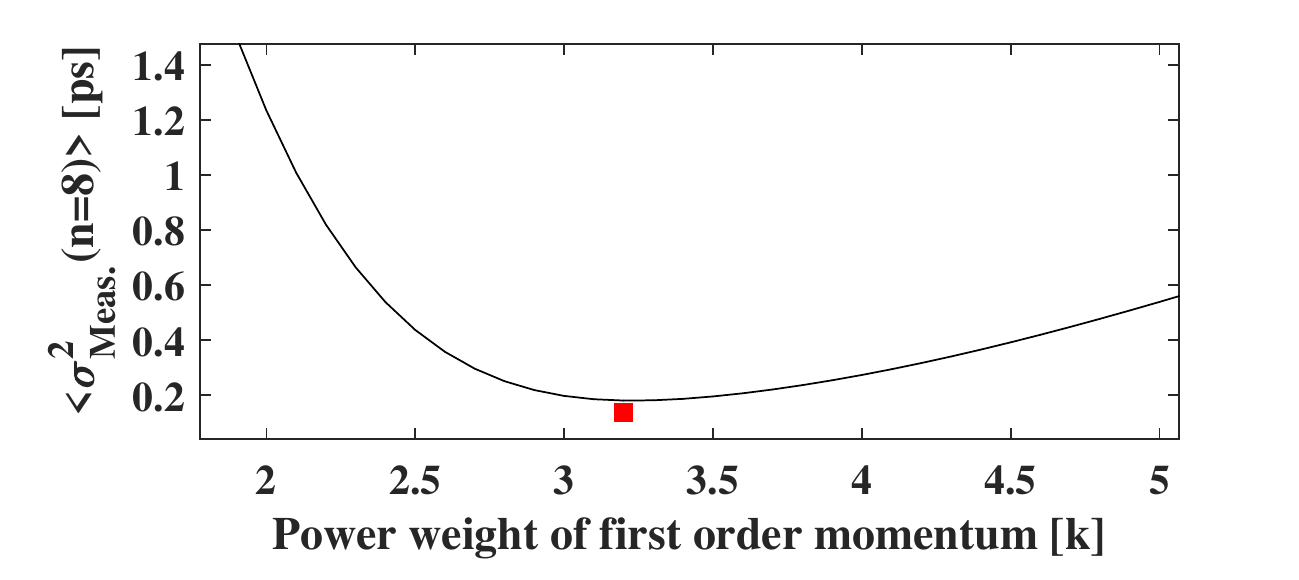}
  \caption{Black line: $<\sigma^2_\text{Meas}(8)>$ versus the power exponent $k$ used in the initial definition of $\tau_0^{Meas}(N)=\int_{N} t.Y(t)^k.dt$. Red square: $<\sigma^2_\text{Meas}(8)>$ after the first pass of super-localization. \label{Afig:1}}
\end{figure}

The initial evaluation of $\tau_0^\text{Meas}(N)$ ($\delta^\text{Meas}(N)$) is given by the computation of the momentum of the timing pulse to the power $k$: $\tau_0^\text{Meas}(N)=\int_{N} t.Y(t)^k.dt$. $k$ is chosen so that our metrics is minimized as seen in \fig{Afig:1}. The resulting  $\tau_0^\text{Meas}(N)$ is shown in \fig{Afig:2}-a). This simple approach allows the determination of the time of passage with an accuracy of 0.18~ps. Using this first estimate of $\tau_0^\text{Meas}(N)$, we compute its corresponding PSF and perform a deconvolution on the timing signal. This first deconvolution reduces the  timing accuracy down to 0.13~ps. The resulting variation of time of passage $\delta^\text{Meas}$ is shown in \fig{Afig:2}-b). If the motion of the soliton molecule round the laser cavity appears now more clearly that in \fig{Afig:2}-a), large pulse to pulse fluctuations remain. We ascribe these fluctuations to the fact that the PSF that we used is not the exact one but a rough approximation of it.

\subsection{Subsequent pass(es) : Self-consistent iteration(s)}
\begin{figure}[htbp]
  \centering
  \includegraphics[width=1\linewidth]{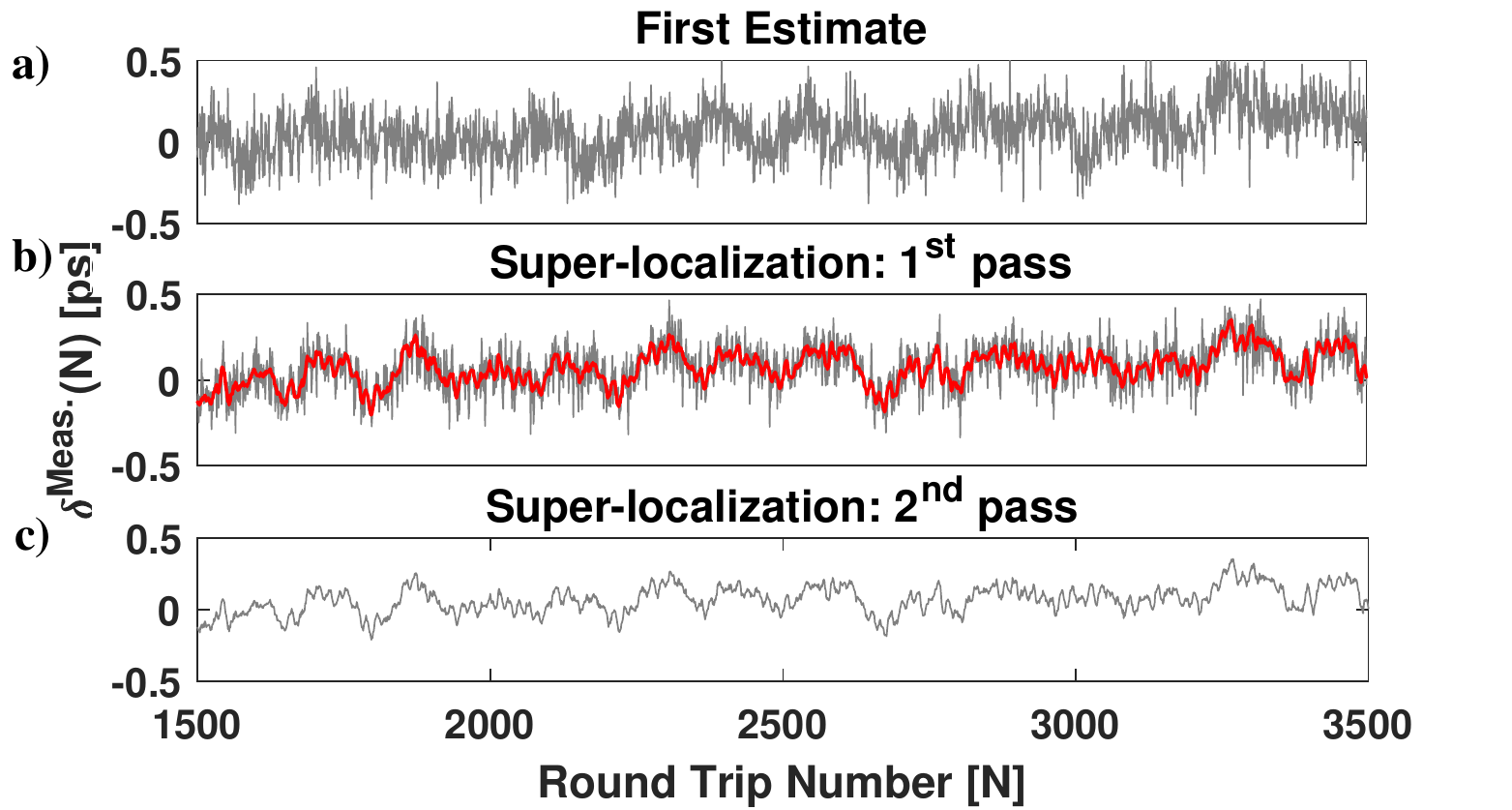}
  \caption{\label{Afig:2} a) $\delta^\text{Meas}(N)$ evaluated using the momentum method with $k=3.2$ ($<\sigma^2_\text{Meas}(8)>=0.18$~ps). b) Super-localization of $\delta^\text{Meas}(N)$: first pass ($<\sigma^2_\text{Meas}(8)>=0.13$~ps). Red line is the smoothed function used as initial guess for the second pass. c) $\delta^\text{Meas}(N)$ obtained after the second - and last - pass ($<\sigma^2_{Meas}(8)>=0.02$~ps).}
\end{figure}

Considering that rapid fluctuations in \fig{Afig:2}-b) are mostly artefacts, we then choose a smoothed version of it (red line \fig{Afig:2}-b) as a new approximation of $\delta^\text{Meas}(N)$. Similarly to what has been done for the first pass, we can now redefine the PSF with better accuracy and deconvolve the timing trace once again. The result of this second pass is shown in \fig{Afig:2}-c). The metrics $<\sigma^2_\text{Meas}(8)>$ tells the numerical timing uncertainty has been reduced down to 0.02~ps.

This procedure can be repeated until convergence. In our case we did not observe any notable improvement of the metrics by performing a third pass. We see that the final $\delta^\text{Meas}(N)$ clearly exhibit distinct temporal features and that most of the spurious peaks seen in \fig{Afig:2}-b) have been removed. The remaining fluctuations ($\approx 100$~fs) are much larger than the metrics ($20$~fs), showing that it corresponds to real experimental jitter. 

\subsection{Hypothesis associated to the deconvolution procedure}
The procedure we just described relies on two major assumptions.

First we assume that the shape of the PSF does not change with the internal state of the molecules. Indeed the molecules are about 6~ps long and they undergo little changes: the total power is stable to a few percent and the internal vibration amplitude of only few hundreds of fs. Consequently, these small changes cannot be recorded by the detection link: electronic residual jitter, 80~GS/s sampling rate, oscilloscope with a 40~GHz electronic bandwidth, optical photodiode with a cut-off around 35~GHz, etc. \textit{A posteriori} this first hypothesis appears to hold. If not, one would have need to short out timing pulses depending of the molecule internal state, and construct one PSF for each internal state. This mitigation plan would nevertheless reduce the number of timing traces used for each PSF, hence reducing the quality of the latter.

The second hypothesis concerns the possibility to define a metrics, hence that the pulse to pulse jitter occurring over one Round-trip is of small amplitude; and that the largest fluctuations are of a much slower dynamics so that it can be simply filtered out through the proper choice of the number of round trips $n$ taken to define the metrics $<\sigma^2_\text{Meas}(n)>$. Considering that the PSF and $\tau_0^\text{Meas}(N)$ are determined using a self-consistent procedure, is it indeed important to define a reliable convergence criteria.

\section{\label{App:B} Second set of soliton molecules}

Here we present the data corresponding to the observation of a second set of soliton-pair molecules using the same experimental setup (Fig. 1-a) and recorded subsequently to the case presented in the main text. 

\begin{figure}[htbp]
  \centering
  \includegraphics[width=1\linewidth]{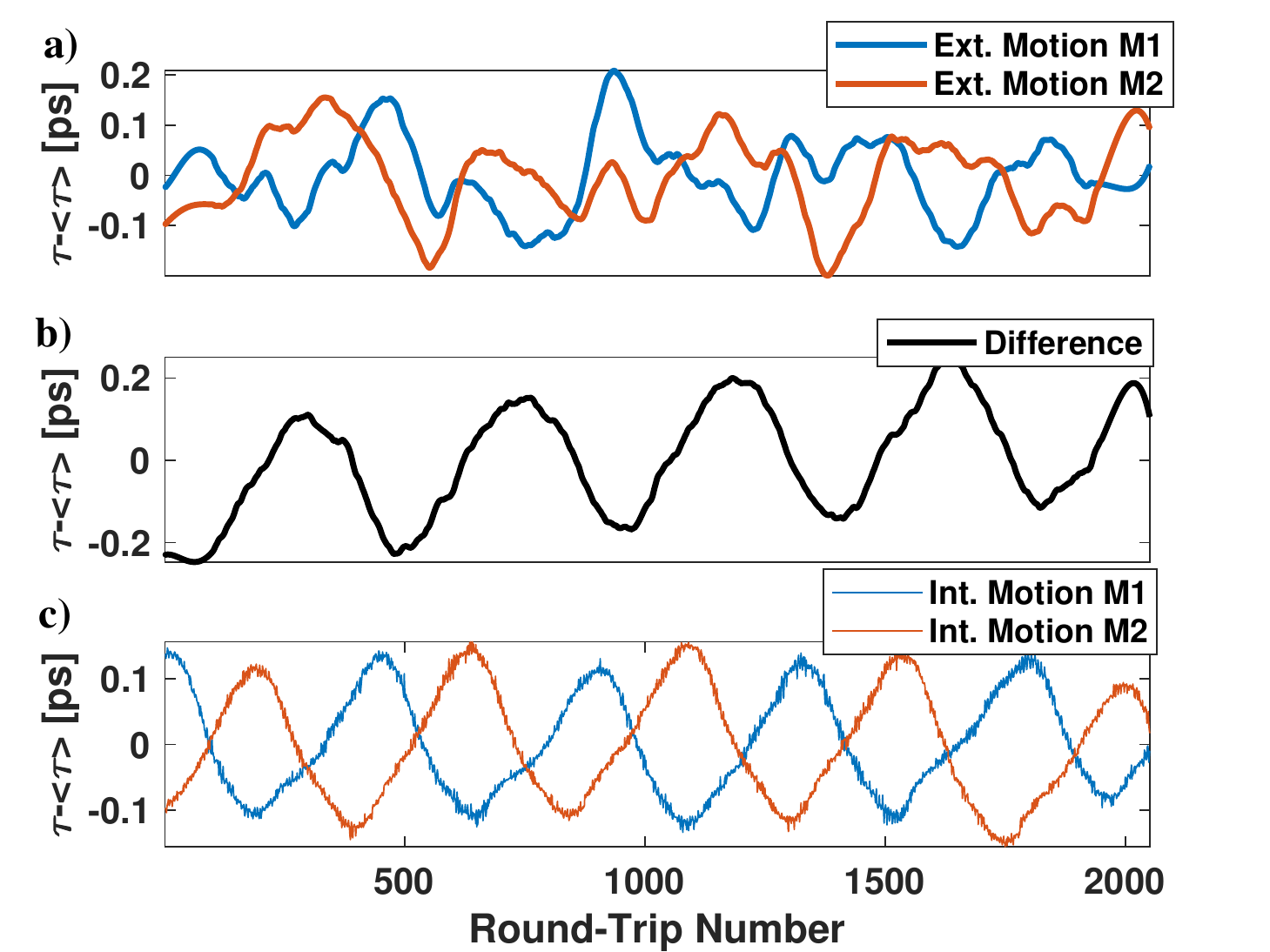}
  \caption{a) Motions of the molecule M1 (blue) and M2 (red). b) Variations of the inter-molecular distance. c) Internal vibration of the molecules (M1=Blue; M2=red). \label{Afig:3}}
\end{figure}

Compared to the first data set, the vibration period is significantly larger (495.6 versus 143.5 RTs), and the two soliton molecules are slightly more compact (temporal extension of 4.6~ps instead of 5.6~ps), see Fig. 7.
A direct consequence of the longer vibration period is that the molecules can be observed during only 5 oscillation periods. Like in the first data set, the separation between the two molecules is close to 1/3 of the cavity round trip time. These similar conditions allow us to compare the two dynamics. 

\begin{figure}[htbp]
  \centering
  \includegraphics[width=1\linewidth]{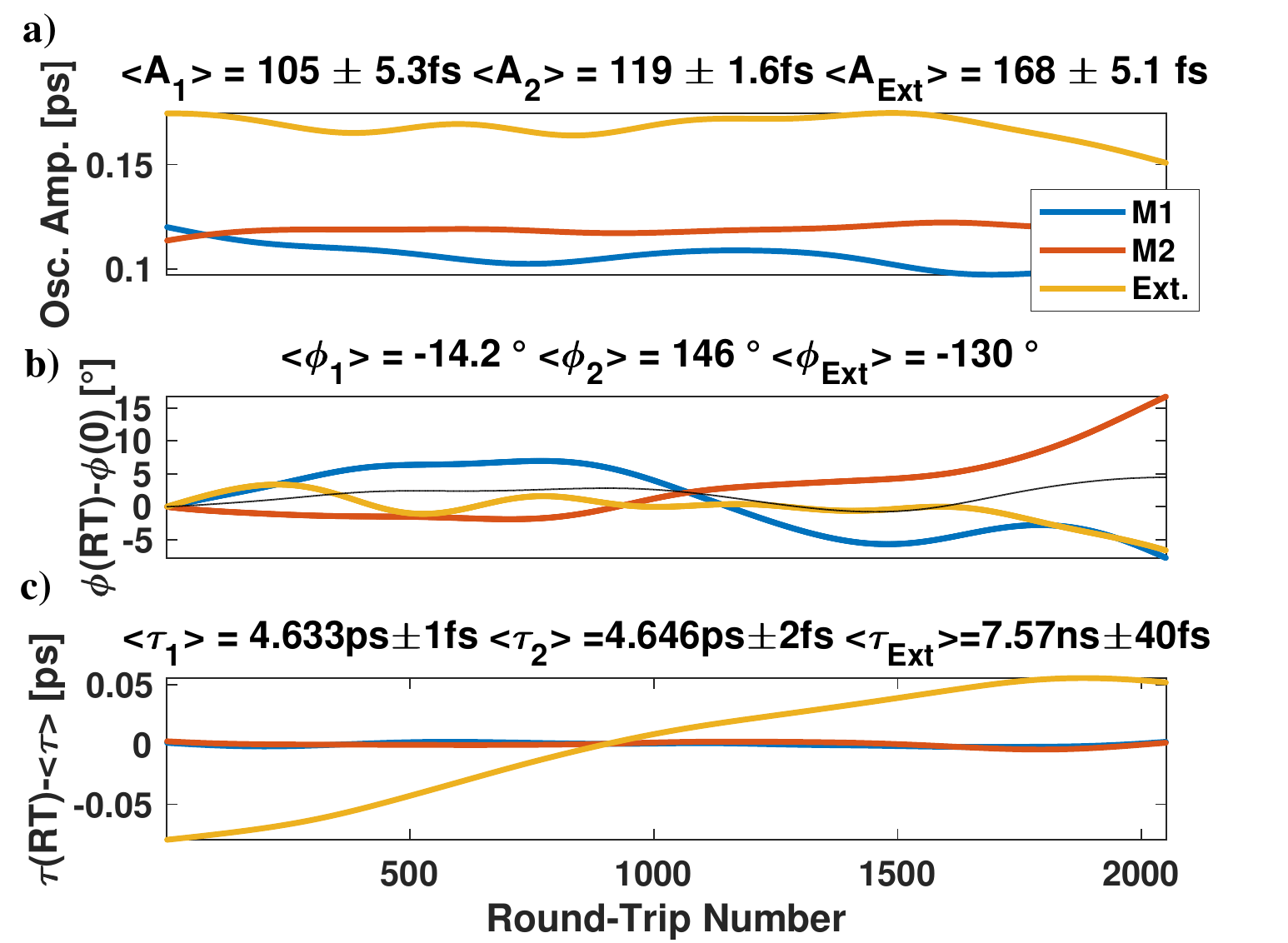}
  \caption{a) Evolution of the instantaneous vibration amplitude for the intra-molecular oscillations (M1=Blue; M2=red) and for the inter-molecular vibration (yellow) b) Same as a) but regarding the instantaneous phase of the vibration patterns. Dashed-black is the results of Eq.(1). c) Evolution of the average soliton separation for M1 (blue) and M2 (red), and of the distance between the two molecules (yellow). \label{Afig:4}}
\end{figure}

By monitoring the evolution of the three oscillators' parameters in Fig. 8 and comparing with \fig{Fig:4}, we can see that the second system is more stable and exhibits less fluctuations. We ascribe this to the overall slower dynamics of the systems, the period of oscillation being 3 times larger. That said, we see in \fig{Fig:4}-c) that the average inter-molecular distance experiences more drift. Note that due to the long oscillation period, the oscillator's are observed here during only 5 cycles, so that their parameters may not be accurately extracted for the few first and last hundreds of round trips (border effect in the numerical treatment). Finally in \fig{Fig:4}-a), we see that the amplitude of motion for the external oscillator can be significantly different from the amplitudes of the internal motions (the oscillators' amplitudes were almost equals for the case presented in the main text).
This second set of data confirms the findings regarding the first set. It also provides some further indications regarding the possible relations between the oscillators' parameters: if the two internal motions are very similar to each other, the external oscillation can have a different amplitude and drifts. The limitation in our present investigations comes from the limited memory of the oscilloscope that prevented us to observe the evolution of vibrating molecules over a larger number of round trips.


\bibliography{Biblio_PiCo.bib}

\end{document}